\documentclass[twocolumn,floats,showpacs,amsmath,amssymb,prb]{revtex4}
                                                                                    
\usepackage{graphicx}
                                                                                   
\begin{document}

\title{Hydrogen and muonium in diamond: A path-integral molecular dynamics 
simulation}
\author{Carlos P. Herrero}
\author{Rafael Ram\'{\i}rez}
\affiliation{Instituto de Ciencia de Materiales,
         Consejo Superior de Investigaciones Cient\'{\i}ficas (CSIC),
         Campus de Cantoblanco, 28049 Madrid, Spain }
\author{Eduardo R. Hern\'andez}
\affiliation{Institut de Ci\`encia de Materials de Barcelona (ICMAB),
         Consejo Superior de Investigaciones Cient\'{\i}ficas (CSIC),
         Campus de Bellaterra, 08193 Barcelona, Spain }
\date{\today}

\begin{abstract}
   Isolated hydrogen, deuterium, and muonium in diamond have been studied 
by path-integral molecular dynamics simulations in the canonical ensemble.
 Finite-temperature properties of these point defects were analyzed in 
the range from 100 to 800 K.
Interatomic interactions were modeled by a tight-binding potential
fitted to density-functional calculations. 
The most stable position for these hydrogenic impurities 
is found at the C--C bond center. 
Vibrational frequencies have been obtained from a linear-response approach,
based on correlations of atom displacements at finite temperatures.
The results show a large anharmonic effect in impurity vibrations at the bond
center site, which hardens the vibrational modes with respect to a harmonic 
approximation.  Zero-point motion causes an appreciable shift of the defect 
level in the electronic gap, as a consequence of electron-phonon interaction.
This defect level goes down by 70 meV when replacing hydrogen by muonium. 
\end{abstract}


\maketitle

\section{Introduction}

In spite of being one of the simplest possible impurities, 
a detailed understanding of the physical properties 
of isolated hydrogen in group-IV materials requires the combination of
advanced experimental and theoretical methods.\cite{pe92,es95}
 Experimental investigations on atomic, interstitial
hydrogen have been so far scarce, since H has turned out to be difficult 
to observe as an isolated impurity in these materials. 
In silicon, electron paramagnetic resonance experiments\cite{go91,ni94} 
indicated that hydrogen is located at a bond-center (BC) site, midway between 
two nearest-neighbor host atoms, in agreement with several theoretical
calculations.\cite{wa89,es95} However, for isolated hydrogen in diamond 
there is little detailed experimental information available in the literature. 
In this context, muonium may be considered as a light pseudoisotope of 
hydrogen, due to the small mass of the muon $\mu^+$ (about 1/9 that of the 
proton). This means that in many respects muonium is expected to behave
similarly to hydrogen, with the appropriate corrections due to zero-point
motion and related effects.
Thus, muon implantation experiments in diamond as well as several theoretical
approaches have shown that this impurity is metastable at a tetrahedral 
interstitial site (the so-called `normal muonium'), and has its lowest energy 
at or around the bond center site (`anomalous muonium').\cite{ho82,es87,es95}

Apart from its basic interest as an isolated impurity, an important property 
of hydrogen in solids is its ability to form complexes and passivate defects. 
This has been extensively studied in the last twenty years for silicon and 
germanium.\cite{pe92,es95} 
More recently, it has been noted that hydrogen can also passivate 
impurities in diamond.\cite{ze99,wa02}
Over the last years, there has been an intense theoretical activity on 
hydrogen-related defects in this material.\cite{mi02,go04,lo04,sq04}
For isolated hydrogen in the neutral charge state, 
most of the calculations found the BC site
as the lowest-energy position for this impurity in diamond.
Some deviations from the BC site were found by Saada 
{\em et al.}\cite{sa00} from a tight-binding (TB) calculation.
Also, for hydrogen in the positive charge state, density-functional theory
calculations have found this impurity to be stable off-axis in a
buckled bond-centered configuration.\cite{go02,go03}

In standard electronic-structure calculations in condensed matter, despite of 
their quantum mechanical character, atomic nuclei are treated as classical 
particles, and therefore typical quantum effects like zero-point vibrations are 
not directly accessible.
These quantum effects can be important for vibrational and electronic
properties of light impurities like hydrogen,
especially at low temperatures. In fact, the importance of taking into 
account the quantum character of proton and muon to study hydrogen-related
point defects in diamond has been emphasized
by Kerridge {\em et al.}\cite{ke04} In particular, quantum tunneling
is relevant to understand the properties of vacancy-hydrogen complexes
in this material.\cite{sh05}  

Finite-temperature properties of hydrogen-related defects in solids
have been studied by {\em ab initio} and tight-binding molecular dynamics (MD)
simulations. In many previous applications of these methods, the atomic 
nuclei were treated as classical particles.\cite{bu91,pa94,be00} 
In order to consider the quantum nature of the nuclei,
the path-integral molecular dynamics (or Monte Carlo) approach has
proved to be very useful.
A remarkable advantage of this method is that all nuclear degrees of
freedom can be quantized in an efficient manner, thus permitting the
inclusion of both quantum and thermal fluctuations in many-body systems
at finite temperatures.  In this way, Monte Carlo or molecular dynamics 
sampling applied to evaluate finite-temperature path integrals allows one to 
carry out quantitative and nonperturbative studies of highly-anharmonic 
effects in solids.\cite{gi88,ce95}

In this paper, the path-integral molecular dynamics (PIMD) method is used 
to investigate the role of
the impurity mass on the properties of hydrogenic point defects.
We consider isolated hydrogen, deuterium (D) and muonium (Mu) in diamond,
in their neutral charge state.
Special attention has been paid to the vibrational properties of these
impurities, by considering anharmonic effects on their quantum dynamics.
 The results of the present calculations show that such anharmonic effects lead
to a significant deviation of the vibrational frequencies of the 
impurities, as compared to a harmonic approximation.  Also, zero-point 
motion causes an appreciable shift of the defect levels in the electronic 
gap, which is a manifestation of the electron-phonon interaction induced
by the hydrogenic impurities.
Path-integral methods analogous to that employed in this work 
have been applied earlier to study hydrogen in metals\cite{gi88,ma95} 
and semiconductors.\cite{ra94,he95,mi98}

 The paper is organized as follows. In Sec.\,II, we describe the
computational method and the models employed in our calculations. 
Our results are presented in Sec.\,III, dealing with the energy of the 
defects, vibrational frequencies, and defect levels in the electronic gap
of diamond.  Sec.\,IV includes a discussion of the results and a summary.

\section{Computational Method}

\subsection{Path-integral molecular dynamics}

In the path-integral formulation of statistical mechanics, the partition
function is evaluated through a discretization of the density matrix
along cyclic paths, composed of a finite number $L$ (Trotter number)
of ``imaginary-time'' steps.\cite{fe72,kl90} In the numerical simulations,
this discretization gives rise to the appearance of $L$ ``beads''
for each quantum particle.  Thus, this method exploits the fact that 
the partition function of a quantum system is formally equivalent to 
that of a classical one, obtained by replacing each quantum particle by a 
ring polymer consisting of $L$ particles, connected by harmonic 
springs.\cite{gi88,ce95}
In many-body problems, the configuration space is usually sampled by
Monte Carlo or molecular dynamics techniques. Here, we have employed the
PIMD method, which has been found to require less computer time resources
in our problem.
Effective algorithms to perform PIMD simulations in the canonical $NVT$
ensemble have been described in detail by Martyna {\em et al.}\cite{ma96} and
Tuckerman.\cite{tu02}
All calculations presented here were carried out in the canonical ensemble,
using an originally developed MD software, 
which enables efficient PIMD simulations on parallel supercomputers.

The calculations have been performed within the adiabatic
(Born-Oppenheimer) approximation, which allows us to define a potential
energy surface for the nuclear coordinates.
As in classical molecular dynamics simulations, an important issue of
the PIMD method is the proper description of interatomic interactions,
which should be as realistic as possible.
To overcome the limitations of effective classical potentials
to reproduce the many-body energy surface, one has to resort to
self-consistent quantum-mechanical methods. Nevertheless, 
true density functional or Hartree-Fock based self-consistent potentials
require computer resources that would restrict enormously the size of
our simulation cell. We have found a reasonable compromise by 
deriving the Born-Oppenheimer surface for the nuclear dynamics from an 
efficient tight-binding effective Hamiltonian, based on density
functional (DF) calculations.\cite{po95}
The capability of tight-binding methods to simulate different properties of
solids and molecules has been reviewed by Goringe {\em et al.}\cite{go97}
We have checked the ability of this DF-TB potential to predict frequencies 
of C--H vibrations. In particular, for a methane molecule it
predicts in a harmonic approximation frequencies of 3100 and 3242 cm$^{-1}$ 
for C--H modes with symmetry $A_1$ and $T_2$, respectively, to be compared 
with experimental frequencies of 2917 and 3019 cm$^{-1}$.\cite{jo93} 
Taking into account the typical anharmonic shift associated to these modes 
(towards lower frequencies), the agreement is satisfactory.
A detailed analysis of vibrational frequencies of hydrocarbon molecules 
derived with the present DF-TB potential, including anharmonicities, can be 
found elsewhere.\cite{lo03,bo01}   

  Simulations were carried out on a $2\times2\times2$ supercell of the 
diamond face-centered cubic cell with periodic boundary conditions,
containing 64 C atoms and one impurity. For comparison, we also carried
out simulations of diamond without impurities, using the same supercell size.
In our calculations, four symmetry independent {\bf $k$} points in the
Brillouin zone of the simulation supercell were employed,
distributed by following the Monkhorst-Pack generation scheme.\cite{mo76} 
We have checked the convergence of the internal energy for some 
selected atom configurations, by considering up to 32 {\bf $k$} points. 
In particular, the internal energy of hydrogen on the tetrahedral T
site and on the BC configuration changes by about 1 meV, respect to
our calculation employing 4 {\bf $k$} points. 
Sampling of the configuration space has been carried out 
at temperatures between 100 and 800 K. 
The simulation-cell parameter employed in our calculations was
taken from experimental data\cite{sk57} and thus changed from
7.1336 \AA\, at 100 K to 7.1424 \AA\, at 800 K.
 For a given temperature, a typical run consisted of $10^4$ MD steps for
system equilibration, followed by $10^5$ steps for the calculation of 
ensemble average properties.

To have a nearly constant precision in the path integral results
at different temperatures, we have taken a Trotter number that
scales as the inverse temperature.
At 300 K we took $L$ = 20 for H and D, and $L$ = 70 for Mu.
For comparison with the results of our PIMD simulations, we have carried
out some classical MD simulations with the same interatomic interaction
(setting $L$ = 1).
Note that for equilibrium properties in the canonical ensemble (i.e.,
mean energy, spatial distribution), the classical limit is equivalent in 
our context to the infinite-mass limit ($m \to \infty$).
The quantum simulations were performed using a staging transformation
for the bead coordinates.
Chains of four Nos\'e-Hoover thermostats with mass $Q = \beta \hbar^2 / 5 L$
were coupled to each degree of freedom to generate the canonical 
ensemble.\cite{tu98}
To integrate the equations of motion, we have used
the reversible reference system propagator algorithm (RESPA), which allows
one to define different time steps for the integration of the fast and slow
degrees of freedom.\cite{ma96} 
The time step $\Delta t$ associated to the calculation of DF-TB forces
was taken in the range
between 0.2 and 0.5 fs, which was found to be appropriate for the
interactions, atomic masses, and temperatures considered here.
For the evolution of the fast dynamical variables, that include the
thermostats and harmonic bead interactions, we used a smaller
time step $\delta t = \Delta t/4$.
We note that for muonium at 200 K (the lowest temperature considered for
this impurity), a simulation run of $10^5$ MD steps requires the calculation 
of forces and energy with the tight-binding code for 
about $10^7$ configurations,
which needs the use of large parallel computers.

\subsection{Calculation of anharmonic vibrational frequencies}

 Important characteristics of impurities in solids are their vibrational
frequencies, which are known to be dependent on the particular site occupied 
by a given impurity in the crystal and on its interactions with the nearby 
hosts atoms.  In our context, the question arises whether the oscillator 
frequencies associated to an impurity located at a given site can be
extracted by assuming the host C atoms fixed in the relaxed geometry. 
This is in fact a method frequently employed to
calculate vibrational frequencies of impurities in solids.
On the other side, when all C atoms are allowed to relax by following 
the impurity motion, the potential energy surface is much flatter 
(smaller energy changes) than when the host atoms are fixed. 
To obtain an approach for the actual vibrational frequencies of the
impurities, one can calculate the eigenvalues of the dynamical matrix
of the whole simulation cell, and obtain the frequencies in the harmonic 
approximation (HA). 
However, for light impurities and atomic configurations
displaying large strains (important relaxations), the anharmonicity can
be appreciable, making the harmonic frequencies meaningless.

To calculate these anharmonic frequencies we will employ a method based 
on the linear response (LR) of the system to vanishingly small forces applied 
on the atomic nuclei.
With this purpose, we consider a LR function, the static isothermal 
susceptibility $\chi^T$, that is readily derived from PIMD simulations of the
equilibrium solid, without dealing explicitly with any external forces in 
the simulation. This approach represents a significant improvement as
compared to a standard harmonic approximation.\cite{ra01}
A sketch of the method is given in the following.

Let us call $\{x_{ip}\}$ the set of $3NL$ Cartesian coordinates of the
beads forming the ring polymers in the simulation cell
($i = 1, \dots, 3N; p = 1, \dots, L$). 
We consider the set $\{X_i\}$ of centroid coordinates, with $X_i$
defined as the mean value of coordinate $i$ over the corresponding polymer:
\begin{equation}
X_i = \frac{1}{L}  \sum_{p=1}^L x_{ip}   \;.
\end{equation}
Then, the linear response of the quantum system to small external forces 
on the atomic nuclei is given by the susceptibility 
tensor ${\chi}^T$, that can be 
defined in terms of centroid coordinates as\cite{ra01}
\begin{equation}
\chi^T_{ij} = \beta \sqrt{m_i m_j} \; \mu_{ij}  \;,
\label{chi_3d}
\end{equation}
where $\beta$ = $(k_B T)^{-1}$, 
$m_i$ is the mass of the nucleus associated to coordinate $i$,
$\mu_{ij} = \langle X_i X_j \rangle - 
             \langle X_i \rangle \langle X_j \rangle$
is the covariance of the centroid coordinates $X_i$ and $X_j$,
and $\langle\dots\rangle$ indicates an ensemble average along a MD run.

The tensor ${\chi}^T$ allows us to derive a LR approximation
to the low-lying excitation energies of the vibrational system,
that is applicable even to highly anharmonic situations.
The LR approximation for the vibrational frequencies reads
\begin{equation}
\omega_{n,LR} = \frac{1}{\sqrt{\Delta_n}}   \;,
\end{equation}
where $\Delta_n$ ($n = 1, \dots, 3N$) are eigenvalues of
$\chi^{T}$, and the LR approximation to the low-lying excitation energy of 
vibrational mode $n$ is given by $\hbar\omega_{n,LR}$. 
More details on the method and illustrations of its ability for predicting
vibrational frequencies of solids and molecules can be found
elsewhere.\cite{ra01,ra02,lo03,ra05}
In connection with the vibrational modes that actually appear in our
calculations, we note that the application of periodic boundary conditions 
is physically equivalent to the only consideration of lattice vibrations at 
the center ({\bf $k$} = {\bf $0$}) of the Brillouin zone of the simulation
cell.\cite{ra05}

\section{Results}

\subsection{Energy}

We first discuss the stable sites for the hydrogenic impurities in 
diamond, as derived from classical calculations at $T = 0$, i.e., point 
nuclei without spatial delocalization. In this respect,
Estle {\em et al.}\cite{es87} noticed the importance of lattice 
relaxation for finding the most stable site for the impurities, and
obtained that interstitial bond-centered hydrogen or muonium is
stable as a result of unusually large lattice relaxation.
This is in fact the case for the DF-TB model employed here. 
After relaxation of the host atoms, the energy surface has minima 
only at two non-equivalent positions. The lowest-energy
position is the BC site, as in several earlier investigations.\cite{go03}
In the relaxed configuration, we find a distance C--H of 1.17 \AA,
which means a backward relaxation of the adjacent C atoms of 0.40 \AA. 
This indicates that the formation of the C--H--C bridge requires a large 
dilation of the C--C bond (a 52\% of the original one). These values agree 
with those found earlier from tight-binding calculations of hydrogen
in diamond.\cite{sa00,ka00}

Another local minimum of the energy surface is found for the impurity 
at the tetrahedral T site, located 1.44 eV above the absolute minimum. 
In this configuration, the relaxation of the nearest lattice atoms is 
much smaller than for the bond-center configuration (0.08 \AA).  
There appears in the literature a rather large dispersion of values for
the relative energy of H on the T site. These values range typically from
0.5 to 2.7 eV.\cite{go03} Then, our result lies in the intermediate region 
of these earlier values, and in particular it is close to that reported by 
Kaukonen {\em et al.}\cite{ka00} from a DF-TB calculation (1.6 eV). 
For the energy barrier from a BC to a T site we find 1.7 eV, a value 
between those obtained from local-density functional theory\cite{go02} 
(1.6 eV) and earlier DF-TB calculations\cite{ka00} ($2.0 \pm 0.1$ eV).
For the migration barrier from T to BC sites we have found 0.3 eV, to
be compared with $0.4 \pm 0.1$ eV obtained in Ref. \onlinecite{ka00}.
We note that the so-called anti-bonding site, between an atom and a T site
along a $\langle 111 \rangle$ direction, has turned out to be from our 
TB calculations a saddle point on the energy surface, contrary to 
Ref. \onlinecite{ka00}, where a local minimum was reported.

We now turn to our simulations at finite temperatures.
The internal energy, $E(V,T)$, at volume $V$ and temperature $T$ can be 
written as:
\begin{equation}
  E(V,T) =  E_{\rm min}(V) + E_{\rm v}(V,T)   \, ,
\label{evt}
\end{equation}
where $E_{\rm min}(V)$ is the potential energy for the classical solid
at $T = 0$, and $E_{\rm v}(V,T)$ is the vibrational energy.
At finite temperatures, $V$ changes with $T$ due to thermal expansion.
The vibrational energy, $E_{\rm v}(V,T)$, depends explicitly on both
$V$ and $T$, and can be obtained by subtracting the energy $E_{\rm min}(V)$
from the internal energy.
In this way, path-integral molecular dynamics simulations allow us to obtain
$E_{\rm v}(V,T)$ for a given volume and temperature.

\begin{figure}
\includegraphics[width=8.5cm]{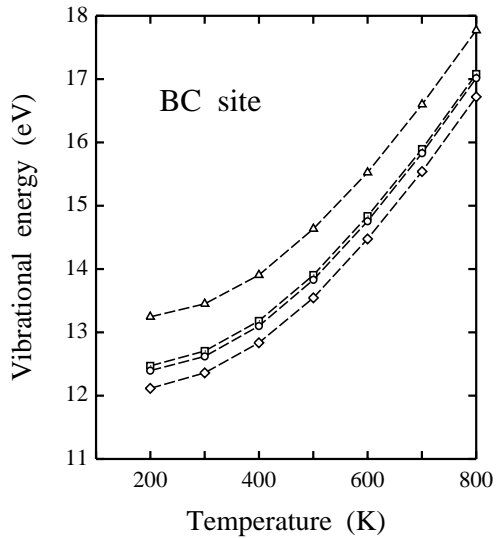}
\vspace{-3.5cm}
\caption{
Temperature dependence of the vibrational energy of the $2\times2\times2$
diamond supercell with one impurity, as derived from PIMD simulations.
Results are shown for deuterium (circles), hydrogen (squares), and muonium
(triangles). For comparison, we also present results for a pure diamond
supercell (diamonds). Dashed lines are guides to the eye.
Error bars are less than the symbol size.
}
\label{f1}
\end{figure}

Shown in Fig.\,\ref{f1} is the temperature dependence of the vibrational 
energy $E_{\rm v}$ per simulation cell for hydrogenic impurities at the BC
site. Symbols indicate results of PIMD simulations: deuterium (circles), 
hydrogen (squares), and muonium (triangles). 
For comparison we also show $E_{\rm v}$ for pure diamond (triangles).  
At 300 K, the vibrational energy of diamond 
amounts to 12.4 eV per simulation cell, i.e., 0.19 eV/atom. 
As expected, $E_{\rm v}$ increases as temperature is raised, and eventually
converges to the classical limit $E_{\rm v}^{cl} = 3 N k_B T$ at high $T$.
When we consider the hydrogenic impurities at the BC site, the vibrational 
energy increases as compared to pure diamond. This increase 
is larger the lower the impurity mass, as a consequence of zero-point motion.

\begin{figure}
\includegraphics[width= 8.5cm]{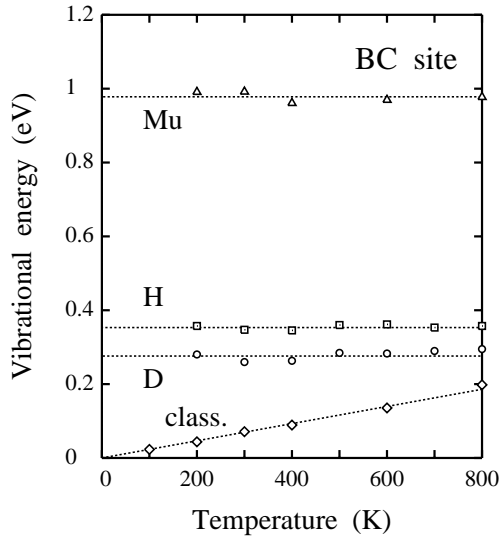}
\vspace{-3.5cm}
\caption{
 Vibrational energy of the hydrogenic defects as a function of temperature.
Results are shown for deuterium (circles), hydrogen (squares), and muonium
(triangles). For comparison, we also present the vibrational energy derived
from classical MD simulations. Error bars are on the order of the symbol
size. Dotted lines are guides to the eye.
}
\label{f2}
\end{figure}

At this point, an interesting characteristic of the different hydrogenic 
defects is their associated vibrational energy. At a given temperature, this
energy is defined as the difference 
$\Delta E_{\rm v}$ = $E_{\rm v}$(64C + Imp) -- $E_{\rm v}$(64C), 
where `Imp' stands for H, D, or muonium.
Note that $\Delta E_{\rm v}$ so defined is not just the vibrational energy 
of a given impurity in the crystal, but it also includes changes in the 
vibrations of the host atoms.   
 In Fig.\,\ref{f2}, the energy $\Delta E_{\rm v}$ is plotted vs 
temperature for the considered impurities at a BC site.
Within the numerical precision of our MD simulations, $\Delta E_{\rm v}$ 
turns out to be constant for the three hydrogenic impurities. 
This means that at temperatures lower than 800 K we find for these
defects, $\Delta E_{\rm v}$ values basically the same as their corresponding 
zero-temperature limit. A slight increase of this energy should
be expected in the temperature region shown in Fig.\,\ref{f2}, especially
for H and D (with vibrational frequencies lower than for Mu), 
due to thermal excitation
of higher vibrational modes. However, it seems that this increase 
is compensated for by a decrease due to lattice expansion, with its associated 
softening of the vibrational modes. In fact, we have carried out a PIMD 
simulation of H at a BC site with the lattice parameter corresponding to 
200 K ($a$ = 3.5668 \AA), and found $\Delta E_{\rm v}$ = 0.40 eV, about 50 meV
higher than the value obtained using the actual lattice parameter at
800 K ($a$ = 3.5712 \AA).
For comparison with our results of PIMD, we also show in Fig.\,\ref{f2} 
results for $\Delta E_{\rm v}$
derived from classical MD simulations (diamonds). These data points follow 
closely the trend expected for a classical three-dimensional harmonic 
oscillator: $\Delta E_{\rm v} = 3 k_B T$.

\begin{center}
\begin{table*}
\vspace{1cm}
\caption{Vibrational energy $\Delta E_{\rm v}$ of hydrogenic defects and 
position of the defect level, $E_I$, at 300 K for impurities at BC 
and T sites, as derived from PIMD simulations. Energy is given in eV, and 
the zero of energy for the electronic levels is taken at the valence-band
top.  Also given are the ratios between vibrational energies for different
impurities. Error bars in $\Delta E_{\rm v}$ are $\pm$ 6 meV for H and D, 
and $\pm$ 10 meV for Mu. The statistical error for $E_I$ is $\pm 2$ meV .
}
\label{tab4}
\vspace{1cm}
\begin{tabular}{ccccc}
 & \multicolumn{2}{c}  {Vibrational energy} &
\multicolumn{2}{c} {Defect level} \\
\colrule
 Impurity     &    BC    &    T   &    BC    &    T  \\
\colrule
 classical  &   0.071  &  0.078  &   2.612  &  3.193  \\
    D       &   0.260  &  0.345  &   2.592  &  3.280    \\
    H       &   0.347  &  0.478  &   2.579  &  3.326   \\
    Mu      &   0.992  &  1.439  &   2.509  &  3.659    \\
\colrule
 $\Delta E_{\rm v}^D$ / $\Delta E_{\rm v}^H$  &   0.75   &  0.72   \\
 $\Delta E_{\rm v}^{Mu}$/$\Delta E_{\rm v}^H$ &   2.86   &  3.01   \\
\hspace{3cm} & \hspace{2cm} & \hspace{2cm} & \hspace{2cm} & \hspace{2cm}
\end{tabular}
\end{table*}
\end{center}

In a one-particle harmonic approximation, the low-temperature vibrational
energies corresponding to deuterium, hydrogen, and muonium scale as
$0.71:1:2.97$, i.e., proportional to $m^{-1/2}$ ($m$, impurity mass).
In Table I we give the vibrational energies of the hydrogenic defects and
ratios between them. At 300 K, $\Delta E_{\rm v}$ values derived from our PIMD
simulations scale as $0.75:1:2.86$, with ratios somewhat different from those
for a harmonic approximation. In fact, we have estimated error bars of
$\pm 0.02$ and $\pm 0.05$ for the ratios
$\Delta E_{\rm v}^D/\Delta E_{\rm v}^H$ and
$\Delta E_{\rm v}^{Mu}/\Delta E_{\rm v}^H$, respectively.
This is an indication of the anharmonicity present
in the defect vibrations, which will be discussed below.
                                                                                     
By using the same procedure, we have calculated the vibrational energy
$\Delta E_{\rm v}$ for impurities at the T site. The results at 300 K are
given in Table I. The resulting energies are larger than the corresponding
ones for the BC configuration. At first sight, this result may seem
surprising, taking into account the constrained geometry of the BC
configuration, which should give a larger frequency for the stretching
vibration along
the C--H--C axis. However, vibrations transverse to this axis are expected
to have much lower frequency, making reasonable a larger $\Delta E_{\rm v}$
for the T defect with threefold degenerate vibrational modes (see below).
For the energy ratios at the T configuration we find $0.72:1:3.01$, which
are compatible within error bars with a harmonic approach.

\begin{figure}
\includegraphics[width= 8.5cm]{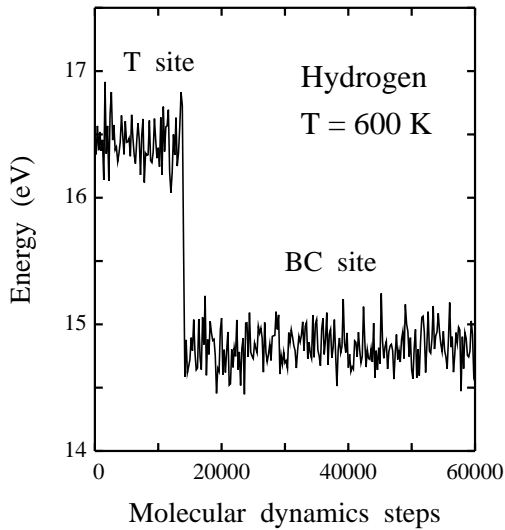}
\vspace{-3.5cm}
\caption{
Internal energy of the simulation cell along a PIMD trajectory at 600 K,
starting with hydrogen on a T site. One observes a drop in the energy
after about 15000 steps, corresponding to a passage of hydrogen from
the T-site region to the BC well.
The zero of energy is taken for a classical system with the impurity
on a BC site at $T$ = 0 (absolute energy minimum).
}
\label{f3}
\end{figure}

A detailed analysis of the temperature dependence of properties of
the impurities at the T site is not feasible due to the metastability
of these defects.
The metastability of hydrogen at the T site is illustrated in
Fig.\,\ref{f3}, where we show the energy of the simulation cell along 
a PIMD run at 600 K. 
This simulation run started with H at a T site, and after about 15000 steps 
we observed the passage of the impurity to a BC site. This is reflected in the 
figure by an energy jump of about 1.6 eV, which corresponds to the 
difference between energy minima plus the difference between vibrational
energies. Similar results were obtained in
other simulations at the same temperature, with hydrogen typically jumping from 
T to BC positions after a number of MD steps in the range from $10^4$ to $10^5$.
However, at 300 K hydrogen remained around the metastable T site during several 
simulation runs of $10^5$ steps, and the same happened for deuterium 
and muonium.
We remember that the time scale of a simulation run (i.e., $\Delta t$ times
the number of MD steps) does not represent a real physical time, since the
dynamics of the beads in the ring polymers does not correspond to the actual 
dynamics of the quantum particles. The fictitious bead dynamics 
is merely used as an efficient way
to calculate equilibrium properties of the system in the canonical ensemble.
Thus, the plot in Fig.\,\ref{f3} illustrates the ability
of PIMD to explore the configuration space, and to drive the system to the
most stable region, once the temperature is high enough.
For comparison, we note that the metastability of muonium at the T site has
been observed by muon spin rotation.\cite{pa88}  
In fact, `isotropic muonium' (at T) converts to `anisotropic muonium' (at
BC) in the range 500--700 K.

\subsection{Vibrational frequencies} 

\begin{figure}
\includegraphics[width= 6cm]{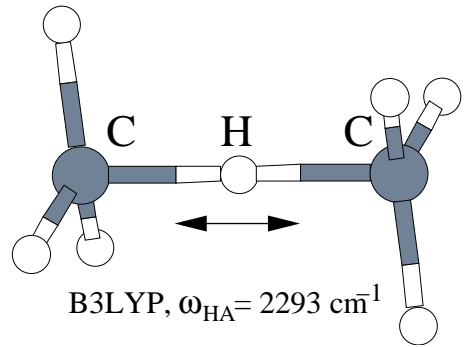}
\vspace{-2cm}
\caption{
Schematic representation of the atomic cluster employed to compare total
energy changes obtained with our tight-binding approach and with
the B3LYP model. $\omega_{HA}$ indicates the vibrational frequency
for hydrogen along the C--C direction, obtained in a single-particle
harmonic approximation in either case.
}
\label{f4}
\end{figure}

\begin{figure}
\vspace{-7cm}
\includegraphics[width=10cm]{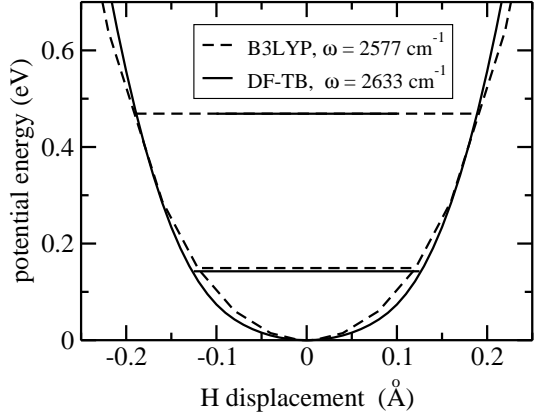}
\vspace{-1.6cm}
\caption{
Potential energy curves for impurity displacements along the direction
parallel to the C--C axis, with fixed carbon atoms, in the cluster shown
in Fig.\,\ref{f4}.  The minimum potential energy corresponds to the impurity 
at the bond-center site.
Horizontal lines represent energy levels for hydrogen, obtained by numerically
solving the one-dimensional Schr\"odinger equation with those potentials.
Results are shown for the tight-binding potential employed here (solid
lines) and for the B3LYP model (dashed lines). The given frequencies
correspond to first excitation energies in both cases.
}
\label{f5}
\end{figure}

As indicated above, the DF-TB potential gives a good description
of vibrational modes in hydrocarbon molecules. 
To check the reliability of this potential to
give vibrational frequencies of hydrogenic impurities in diamond, 
and to assess their anharmonicity, we have carried
out calculations for a simple atomic cluster (see Fig.\,\ref{f4}).
In this cluster, the impurity  occupies a position similar to a BC site 
in the diamond lattice, midway between two carbon atoms with a distance 
$d$(C,H) = 1.17 \AA, as that found in diamond with the DF-TB
potential. In this atomic cluster, we have
calculated total energies for hydrogen displacements along the C--C axis,
with both our TB Hamiltonian and the B3LYP density-functional theory
(with the cc-pDVZ basis set as implemented in the Gaussian 98 package). 
\cite{ga98}
Results obtained by both methods are shown in Fig.\,\ref{f5}, where 
the zero of energy corresponds to the BC position in either case. 
The continuous and dashed
lines correspond to DF-TB and B3LYP models, respectively.
The curvature of the solid line at the minimum is smaller than that of
the dashed line, and thus the harmonic approach gives for the DF-TB method
a vibrational frequency (1991 cm$^{-1}$ for H) lower, but of the same order, 
than that of the B3LYP potential (2293 cm$^{-1}$ for H).
We emphasize that these energy curves do not correspond to hydrogenic
impurities in bulk diamond, but they help to analyze the reliability of
the DF-TB potential and to illustrate qualitatively the vibrational
potential of these impurities in the solid.

In Fig.\,\ref{f5} we also present the two lowest energy levels for hydrogen,
obtained by numerically solving the one-dimensional Schr\"odinger equation 
with each interaction model. 
Now the ground state for the B3LYP potential is higher than that of the 
DF-TB model.  The first excited states
yielded by both procedures are very close to one another, and in fact are
indistinguishable on the scale of Fig.\,\ref{f5}. We then find that the 
excitation energy
$\hbar (\omega_1 - \omega_0)$ corresponds to 2633 and 2577 cm$^{-1}$ 
for DF-TB and B3LYP, respectively.
In both cases we find excitation energies larger than in the harmonic
approximation, i.e., anharmonicity hardens the stretching vibration, as 
expected for an impurity in a highly-confined geometry. 
The excitation energies obtained by both methods are similar,
with a relative difference between them smaller than 3\%, which gives us 
confidence in the reliability of the DF-TB method to analyze vibrational 
properties of hydrogenic point defects in diamond.

\begin{figure}
\includegraphics[width= 10cm]{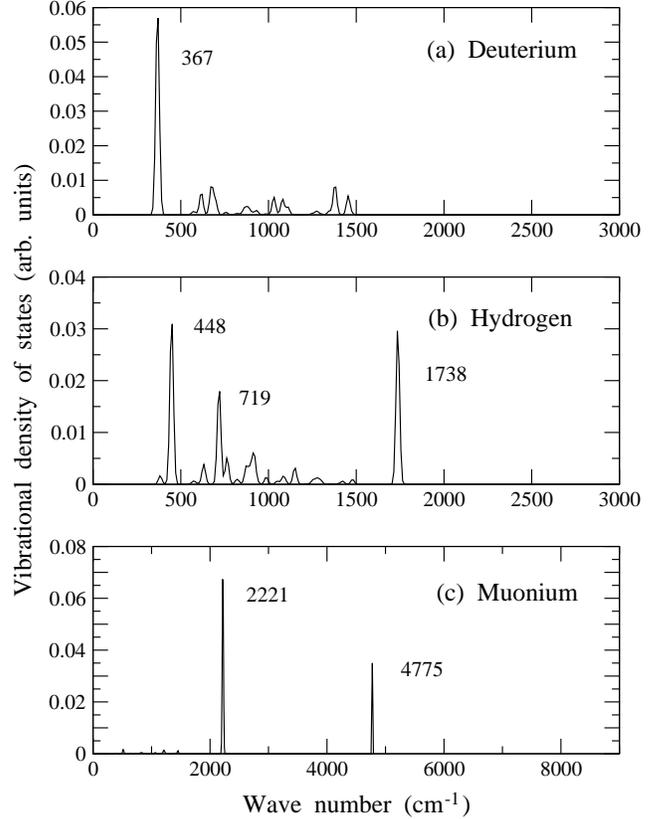}
\vspace{-3.3cm}
\caption{
Vibrational density of states for impurities at a BC site,
as derived from the dynamical matrix of the simulation cell
in a harmonic approximation.  Each panel corresponds to an impurity:
(a) deuterium, (b) hydrogen, (c) muonium.
Labels indicate wave numbers in cm$^{-1}$.
For the sake of clarity, discrete modes have been broadened to Gaussian
profiles with a standard deviation of 10 cm$^{-1}$.
Note the different horizontal scale in panel (c).
}
\label{f6}
\end{figure}

We now go back to the study of hydrogenic impurities in bulk diamond.
Before analyzing the anharmonic modes given by the linear response
procedure described in Sec. II.B, we will present the vibrational
frequencies of the impurities derived from a harmonic approximation.
This will allow us to assess the importance of anharmonicity
for the impurity vibrations.
By calculating the dynamical matrix, we find the vibrational density of 
states (VDOS) corresponding
to the whole simulation cell (64 C atoms plus one impurity).
We then use the amplitude of the displacement of the impurity 
in each vibrational mode, to obtain its partial VDOS. 
This is shown in Fig.\,\ref{f6} for (a) deuterium, (b) hydrogen, and 
(c) muonium at the BC site.
The results can be most clearly understood for the case of muonium,
where one observes basically two peaks at 2221 and 4775 cm$^{-1}$,
well separated from the region of crystal vibrations
(up to about 1500 cm$^{-1}$). 
Due to the low mass of Mu, these vibrations are almost totally decoupled
from the host-atom vibrations, and are associated to motion perpendicular
(bending, twofold degenerate) and parallel (stretching) to the C--C axis, 
respectively.
This is in fact the situation expected for an impurity at a BC site, assuming 
fixed host atoms.  The partial VDOS is, however, more complicated for H and D,
which participate in modes that are not totally
localized at the impurity atom or at the impurity plus its nearest C neighbors. 
For H, we find only one mode with frequency larger
than the lattice modes, corresponding to the stretching vibration along the 
C--H--C bond ($\omega_{\|,HA} =$ 1738 cm$^{-1}$). 
We obtain two other peaks at 448 and 719 cm$^{-1}$ 
with large participation of hydrogen motion. They are assigned to transverse 
vibrations, perpendicular to the C--H--C axis, and include important 
participation of nearby C atoms. 
For deuterium, the most prominent feature is a large peak at 367 cm$^{-1}$,
corresponding to bending vibrations.
In this case, the stretching vibration is highly coupled to crystal modes,
and does not appear as a prominent peak in the VDOS. 

\begin{figure}
\includegraphics[width=10cm]{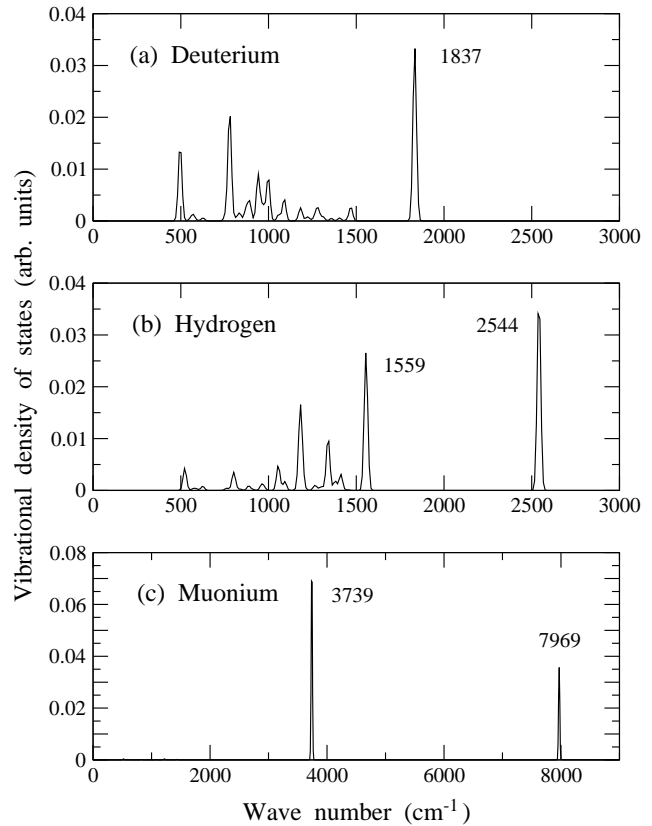}
\vspace{-2cm}
\caption{
Vibrational density of states for impurities at a BC site and
$T$ = 300 K, as derived from a linear-response analysis
of PIMD results. Each panel corresponds to an impurity:
(a) deuterium, (b) hydrogen, (c) muonium.
Labels indicate wave numbers in cm$^{-1}$.
For clarity, discrete modes are presented with a Gaussian profile
with a standard deviation of 10 cm$^{-1}$.
Note the different horizontal scale in panel (c).
}
\label{f7}
\end{figure}

Next, we apply the linear-response procedure to
obtain the (anharmonic) vibrational frequencies of the impurities. 
Again, we obtain the VDOS of the whole simulation cell and project
from it the contribution of the considered impurity.
This partial VDOS is shown in Fig.\,\ref{f7} for the hydrogenic impurities
studied here at 300 K.
As before, we first comment on the results for muonium in panel (c).
We observe two distinct peaks, as for the HA, but now at much higher
frequencies. This resembles the effect of anharmonicity observed for the
atomic cluster considered above. For muonium at a BC site in diamond,
both vibrations parallel and perpendicular to the C--Mu--C axis harden 
strongly due to anharmonicity.
Something similar happens for hydrogen and deuterium, as noticed when one
compares the corresponding panels in Figs.\,\ref{f6} and \ref{f7}.
For hydrogen, an interesting feature is the appearance of a second peak
at 1559 cm$^{-1}$, above but close to the largest frequency in bulk diamond, 
corresponding to a bending mode [Fig.\,\ref{f7}(b)]. 
For deuterium, we obtain a well-resolved stretching mode at 1837 cm$^{-1}$, 
far in frequency from the lattice vibrations.
For the frequencies derived from the LR method, we estimate error bars of
20 cm$^{-1}$, due to the statistical noise present in the susceptibility 
tensor $\chi^T$ derived from the PIMD simulations.
For the different impurities, the highest-frequency mode $\omega_{\|,LR}$ 
scales as $0.72:1:3.13$, to be compared with the ratio $0.71:1:2.97$ expected 
for one-particle harmonic oscillators. 
The stretching frequency derived for H from our LR calculations 
($\omega_{\|,LR} =$ 2544 cm$^{-1}$) is lower than that obtained for this 
impurity by Goss {\em et al.}\cite{go02,go03} from DF calculations in the 
harmonic approximation (2919 cm$^{-1}$). This difference can be explained,
at least in part, by the difference in the C--H distance obtained by both
methods: 1.17 \AA \, for our TB potential vs 1.13 \AA \, for the DF
calculations\cite{go02} (the larger the distance, the lower is expected 
to be this frequency). We also note that DF calculations predict a 
frequency of 2456 cm$^{-1}$ for positively charged hydrogen in a
buckled bond-centered configuration.\cite{go02}

We have repeated the same LR procedure to calculate the vibrational
frequencies of hydrogenic impurities at the T site. In this case,
we find for each impurity a threefold degenerate mode, in agreement with
the tetrahedral point symmetry of this site. Coupling with lattice
vibrational modes is negligible. We find the frequencies 2088, 2818,
and 7284 cm$^{-1}$ for D, H, and Mu, respectively, which translates to a
frequency ratio 0.74:1:2.58. 
For impurities at the T site, anharmonicity causes a softening of the modes.
For example, for hydrogen we find in the HA a frequency of 3099 cm$^{-1}$,
more than 200 cm$^{-1}$ larger than that found with the LR method.
This softening is still more important for muonium, due to its larger
zero-point vibration, and is the origin of the small ratio 
$\omega_{||,LR}^{Mu}/\omega_{||,LR}^H = 2.58$ (to be compared with a ratio of 
2.97 expected for harmonic oscillators).
We note that anharmonicity affects the mode frequencies in
opposite ways at the BC and T sites. While the vibrational modes
associated to the impurities at the BC site suffer a strong shift towards
higher frequencies, for the impurities at the tetrahedral site the shift
is towards lower ones. This no doubt is related to the different
geometries in both configurations: at the BC site the impurity is in a
much more confined environment, with a strongly directional character
(along the bond), while at the T site the impurity is less constrained.

\subsection{Defect levels in the electronic gap} 

The DF-TB method employed here allows us to study the influence of
zero-point motion on the one-electron levels. For a classical hydrogen-like
impurity at a BC site, we find a defect level located in the 
electronic gap 2.61 eV above the top of the valence band.
This result is close to the donor level found by Goss {\em et al.}
from local-density-functional calculations.\cite{go02} 
Also, Saada {\em et al.} found a defect level at $\sim$ 0.5 eV above
the middle of the energy gap for the minimum-energy configuration, which
in their case corresponded to the impurity on off-BC sites.\cite{sa00}  

Defect levels associated to the presence of the impurity in the crystal
are expected to change as a consequence of electron-phonon interaction.  
In fact, effects of this kind of interactions on the electronic structure 
have been experimentally investigated by measuring the temperature
dependence of the optical spectra of solids.\cite{ca05a} These effects
appear even at low $T$, due to zero-point motion, and can be large in 
some cases. For example, in diamond the direct electronic gap (at the 
$\Gamma$ point) is reduced by about 0.6 eV with respect to that calculated 
when zero-point vibrations are neglected.\cite{ca05a} 
Such studies have been complemented in some cases by observing the 
dependence of the spectra on isotopic mass, whenever different stable 
isotopes are available.\cite{zo92,ca05b}

\begin{figure}
\includegraphics[width= 8.5cm]{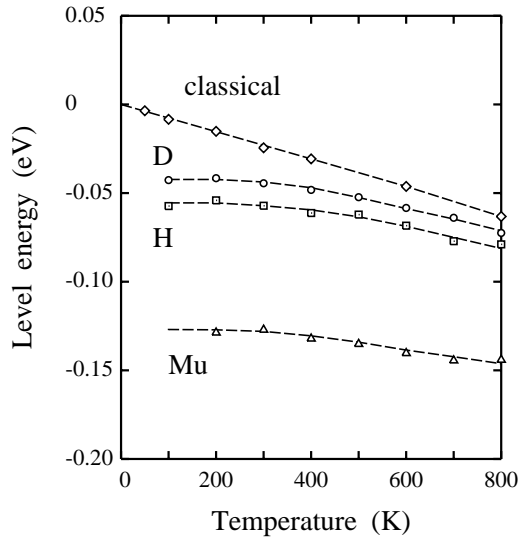}
\vspace{-3.5cm}
\caption{
Temperature dependence of the defect-level energy for deuterium (circles),
hydrogen (squares), and muonium (triangles) at a bond center.
For comparison, we also present results for a classical simulation (diamonds).
The zero of energy corresponds to the level of a classical impurity at $T$ = 0.
Error bars are on the order of the symbol size. Dashed lines are guides
to the eye.
}
\label{f8}
\end{figure}

Similar effects of quantum atomic vibrations should also appear
in the defect levels induced by the hydrogenic impurities studied here.
Shown in Fig.\,\ref{f8} is the temperature dependence of the defect
level energy for H, D, and Mu at a BC site in diamond, as derived 
from our PIMD simulations.  
For comparison, we also present the energy level $E_I$ derived from classical 
MD simulations. In this Figure, 
the zero of energy corresponds to the classical limit at $T$ = 0.
The most remarkable fact that one observes in this
plot is an appreciable shift of the energy level as the impurity 
mass changes. In particular, for the impurity at the BC site, 
the level goes down as the mass is lowered. 
At 300 K we find a decrease in $E_I$ of 13 meV when passing from D to H,
and a further decrease of 70 meV when comparing muonium with deuterium.
We note also that the level in the case of deuterium is 20 meV lower
than that corresponding to a classical impurity (limit of infinite
mass) at 300 K.   In all cases, the statistical error
due to finite sampling of the canonical ensemble is of about 2 meV.

For a given impurity, the defect level goes down as the temperature is raised.
In fact, for H and D we find a decrease in the level energy of about
30 meV when $T$ is increased from 100 to 800 K.
For muonium, this temperature-induced change amounts to about 20 meV.
For the classical impurity this level shift is more important, resulting
to be 63 meV.

The position of the defect level for the different impurities at 300 K is 
given in Table II, where we also present the energy of the level for 
impurities at the tetrahedral T site.  In this case, the classical result at 
300 K is found to be 0.58 eV above the defect level for the impurity at a BC
site.  

For hydrogenic atoms at site T, 
the level goes up as the impurity mass is lowered, contrary
to the case of the bond-center configuration.  Also, the level for the
T configuration is more sensitive to the impurity mass, since it changes by
0.47 eV from a classical impurity to muonium, vs a decrease of 0.10 eV
for the BC configuration.
 
We note that, while the absolute position of the defect level can depend
on the model employed, its relative change induced by temperature
and isotopic mass is expected to be quite reliable, because 
in our analysis both electronic and nuclear dynamics are treated at the level 
of quantum mechanics, which permits a more realistic description of the 
electron-phonon interaction in this system.

\section{Discussion}

In Sec. III we have presented results of our PIMD simulations for H, D,
and Mu in diamond. 
The main advantage of this kind of simulations is the possibility
of calculating defect energies at finite temperatures, with the
inclusion of a full quantization of host-atom motions, which are not 
easy to be accounted for in fixed-lattice calculations. 
Isotope effects can be readily explored, since the impurity mass
appears as an input parameter in the calculations. This includes the
consideration of zero-point motion, which together with anharmonicity
causes appreciable non-trivial effects.
 Our results indicate that hydrogenic impurities at the bond-center site 
cannot be accurately described as a particle moving in a harmonic potential. 
Even if anharmonicities of the interatomic potential are taken into account,
a single-particle approximation does not give reliable results for the 
impurity complex at finite temperatures. It is then necessary to treat
the defect as a many-body problem with anharmonic interactions. 
                                                                                    
The most prominent feature in the VDOS of the hydrogenic impurities at a
BC site is a peak corresponding to the stretching vibration, with a
frequency $\omega_{\|}$ higher than the lattice vibrations.
This stretching mode is infrared-active, but to our knowledge has not 
been detected yet. On the theoretical side, there are not many calculations 
in the literature predicting the frequency of this mode.
Goss {\em et al.}\cite{go02} employed an HA from local-density-functional
calculations and found for hydrogen at a BC site (in the neutral charge 
state) a stretching frequency $\omega_\|$ = 2919 cm$^{-1}$, somewhat higher 
than our result using LR calculations ($\omega_{\|,LR}$ = 2544 cm$^{-1}$).

From PIMD simulations we have calculated in Sec. III.A the defect energy
$\Delta E_{\rm v}$ for the different hydrogenic impurities. Also, by using
the LR procedure we have obtained an approach to the vibrational frequencies
of the impurities, which is more realistic than a pure harmonic
approximation.  These frequencies can now be used to estimate a zero-point
energy for the impurities in diamond. In the case of muonium, the modes
are almost totally localized on the impurity, and the zero-point vibrational
energy of the defect is well approximated by
$\Delta E_{{\rm v},LR}^0 = \hbar (\omega_{\bot,LR} + \omega_{\|,LR} /2)$.
This gives $\Delta E_{{\rm v},LR}^0$ = 0.96 eV, close to the ground state energy
of the defect presented in Fig.\,\ref{f1} ($0.99 \pm 0.01$ eV). 
Using the frequencies $\omega_{\bot,HA}$ and $\omega_{\|,HA}$ 
derived from the HA, one finds $\Delta E_{{\rm v},HA}^0$ = 0.57 eV, 
much lower than that obtained directly from PIMD simulations.
Note that we have used here an expression for $E_0$ which is only rigorously
valid for harmonic oscillators, but it nevertheless gives 
a good approximation to the
zero-point energy of the defect when one introduces the frequencies derived
from the LR method. In this sense, these LR frequencies may be
considered as renormalized
harmonic frequencies, which incorporate in a nonperturbative way
anharmonicities of the interatomic interactions, as seen by the atoms in
their quantum motion.

  The zero-point energy of the defect requires a comment in the sense
that, in the case of muonium, it could help to cross the  adiabatic
barrier for impurity diffusion.  Nevertheless, we find that muonium is 
confined in the BC region, especially at low temperatures.
Impurity migration to other regions of the crystal requires important
relaxations of the neighboring C atoms, which are not probable at
low temperatures. This picture is similar to that
described in the literature as ``opening of a door'', \cite{bo94}
which favors impurity diffusion.

  An analysis of hydrogen diffusion in diamond is out of the scope 
of this paper.  As noted above, actual diffusion coefficients are not 
directly accessible with the kind of MD simulations employed here, 
since the time scale in the simulations is not readily connected to the
real one.  In connection with this, PIMD simulations can be applied 
to study quantum diffusion (including tunneling) of hydrogen in pure 
and doped diamond, by calculating free-energy barriers in a way similar 
to that employed earlier to study H diffusion in metals\cite{gi88} and 
semiconductors.\cite{he97} 
Even though the rate of H tunneling in diamond is not expected to be high
at low $T$, due to the large lattice relaxation associated to the 
impurity on a BC site, thermally activated tunneling may be possible,
as observed for hydrogen in silicon.\cite{he97}  
This point will require further research.

Theoretical techniques to calculate the electronic band structure of solids
have been improving their precision for many years. For various purposes, the
accuracy currently achieved by these methods is excellent, when comparing
their predictions with experimental data. However, zero-point motion
is a significant factor limiting the accuracy of state-of-the-art
techniques to predict energy bands.
The same happens for defect levels caused by impurities in solids, 
since their energy may change appreciably as the impurity mass is varied. 
This effect has been illustrated here by shifts in the energy of the
levels due to hydrogenic impurities in BC or T sites in diamond.
For muonium at the T site, we have found a level shift of 0.47 eV (upwards), 
on the order of the renormalization of the diamond gap due to
zero-point motion ($\sim 0.6$ eV). For the BC site, the corresponding shift 
for Mu was 0.10 eV and for H we found 0.03 eV (both downwards).
Thus the magnitude and direction of these shifts 
depend markedly on the mass and position of the impurity in the solid. 
This behavior can be rationalized by considering the results  
of perturbation theory for the electron-phonon interaction in 
the limiting case of a harmonic oscillator.\cite{ca05a,ca05b}
The shift of an electronic level is expected to be proportional to the
square of the mode amplitude, i.e., proportional to $m^{-1/2}$ in the 
low-temperature limit and to $T$ at high temperatures.  The extrapolation 
to zero temperature of the defect levels for D, H, and Mu at a BC site
(see Fig.\,\ref{f8}) gives shifts scaling as $0.72:1:2.24$, to be compared to 
the ratio $0.71:1:2.97$ expected for a harmonic impurity.
The energy shift for Mu deviates appreciably from the harmonic 
expectation, as a consequence of the larger anharmonicity of  
this impurity center.

In summary, the PIMD method has turned out to be well-suited to study
finite-temperature equilibrium properties of light impurities 
in diamond.   This has allowed us to notice the importance of 
anharmonicity in order to give a realistic description of the 
properties of these point defects. 
This anharmonicity shows up in the vibrational modes of the impurities,
causing important shifts respect to the harmonic expectancy.
Also, zero-point motion introduces an appreciable shift in the defect 
levels, which depends on the impurity mass. 
This type of analysis offers a promising way for studying other challenging 
effects of light impurities in diamond, such as quantum diffusion.

\begin{acknowledgments}
The calculations presented here were performed at the Barcelona
Supercomputing Center (BSC-CNS).  
This work was supported by CICYT (Spain) through Grant
No. BFM2003-03372-C03-03. ERH thanks DURSI (regional government of Catalonia)
for funding through project 2005SGR683.
\end{acknowledgments}

\bibliographystyle{apsrev}

\end{document}